# Application of Event-Triggered Sliding Mode Control of 2-DOF Humanoid's Lower-Limb Powered by Series Elastic Actuator

Anh Khoa Lanh Luu[3], Van Tu Duong[1,2,3], Huy Hung Nguyen[3,4] and Tan Tien Nguyen[1,2,3,*]

[1] Faculty of Mechanical Engineering, Ho Chi Minh City University of Technology (HCMUT), 268 Ly Thuong Kiet, District 10, Hochiminh City, Vietnam.
[2] Vietnam National University Ho Chi Minh City (VNU-HCM), Linh Trung Ward, Thu Duc District, Hochiminh City, Vietnam.
[3] National Key Laboratory of Digital Control and System Engineering, Ho Chi Minh City University of Technology, VNU-HCM, Hochiminh City, Vietnam.
[4] Faculty of Electronics and Telecommunication, Saigon University, Hochiminh City, Vietnam.
*Corresponding author's email: nttien@hcmut.edu.vn

**Abstract.** This paper proposes an event-triggered sliding mode control (SMC) scheme combined with a backstepping algorithm for control of 2-DOF humanoid's lower-limb powered by Series elastic actuator (SEA). First, the modelling process for the lower-limb system is implemented by using the Euler-Lagrange theory. With the obtained dynamical equations of lower-limb, the model of the SEA is achieved in both mechanical and electrical perspectives. Then, the event-triggered SMC approach is utilized to ensure the system's stability and eliminate the effect of bounded external disturbance. Next, some assumptions and designed thresholds for the tracking error are given, together with the proof for the convergence of the inter-event time. The backstepping algorithm is applied in the end to determine the needed input control voltage signal. Finally, the results of this research are demonstrated through some simulations in order to prove the efficiency and appropriation of this method.

**Keywords:** Event-triggered Sliding Mode Control, Series Elastic Actuator, Humanoid Lower-limb.

## 1  Introduction

Conventionally, the control input signal from the controller applying to the plant is updated at the sampling instant only and kept unchanged between two consecutive sampling instants. This feature is usually called the *time-triggered control* strategy, which is widely applied in most simple systems due to its massive advantage of simplicity for both programming and system analysis [1]. However, this approach has a significant drawback about the system's sensitivity to the output state when the control input is triggered periodically regardless of the tracking error and the changing rate of the output variables. Besides, from the communication perspective, the gradual generation and transmission of a control input signal in every instant are usually considered a waste of resources and unproductive, which can also lead to the delay phenomenon for the whole controlled system.

For those reasons, an alternative approach, called *event-triggered control*, has been proposed and analyzed in many research to deal with the downsides of the traditional strategy [2], [3]. By considering stabilization conditions and some desired error limits, the control input's triggering instant is now heavily relied on the system's output states. This approach has shown its remarkable effectiveness in the steady-state of the controlled system, which made it widely accepted and applied worldwide in many scientific research and industrial fields. Recently, event-triggered strategy has gradually been used in spacecraft for attitude control and flying stabilization, where remote control and external disturbance are always the critical problems. In [4], [5], the event-triggered control scheme was applied to reduce the wireless communication pressure between the controller and the aircraft's actuators as well as to give more resources for other vital sensors' feedback signals. Besides, this approach has also proved its appropriation for the trajectory tracking problem of autonomous vehicles [6], [7], and industrial robots [8], [9]. In those applications, reducing the control signal commutation frequency impacted the decrement of the data broadcasting between the controller and the plant. Also, the updates of the control signal could be suitably tuned for a specific application so as to fulfil particular control requirements.

In this research, the event-triggered SMC is utilized to control the motion of the 2-DOFs humanoid's lower-limb powered by a Series elastic actuator (SEA) in the swing phase. The SEAs have been selected to power the humanoid's joints because of various prominent advantages compared to the traditional stiff actuators [10]. In detail, the SEA can absorb high-frequency shock from the load to protect the physical system, store and release energy thanks to the elastic element, and turn the force control problem into the position control problem with high fidelity. Many reputable functional humanoids have successfully operated with the integration of the SEAs [11], [12], which proved their efficiency in the robot locomotion field.

In this paper, the Euler-Lagrange model of the lower-limb system is firstly given together with the overall dynamical equations. Then, the SEA modelling is briefly demonstrated by referring to our previous research [13], [14]. Next, the event-triggered SMC algorithm is utilized to design the controller for the obtained model. Besides, the backstepping technique is applied to determine the desired control input voltage. Finally, some simulations are implemented to evaluate the lower-limb system's response quality and the effectiveness of the event-triggered approach.

## 2  System Modeling

### 2.1  Lower-limb modelling

The lower-limb system is demonstrated as a double compound pendulum system with two links powered by two SEAs (assumed as massless compared to the links). The points of $M_1$ and $M_2$ are the centre of mass's (COM) positions at each link (as shown in **Fig. 1**). By making translational movements, the SEAs generate rotational motion of the links about their joints with rotational angles of $\theta_1$ and $\theta_2$. The angles of $\alpha_1, \alpha_2, \sigma_1, \sigma_2$ are constant and defined related to the dimensions of each link.

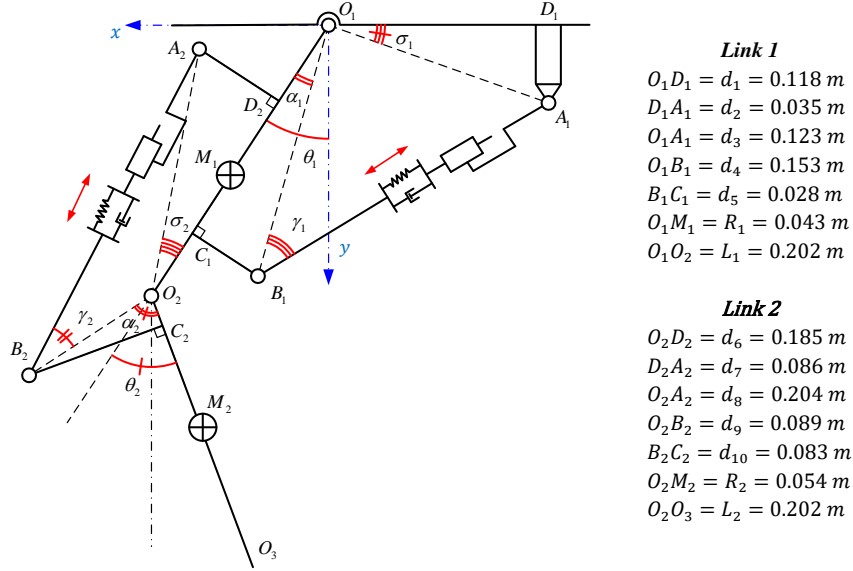

**Link 1**
$O_1D_1 = d_1 = 0.118\ m$
$D_1A_1 = d_2 = 0.035\ m$
$O_1A_1 = d_3 = 0.123\ m$
$O_1B_1 = d_4 = 0.153\ m$
$B_1C_1 = d_5 = 0.028\ m$
$O_1M_1 = R_1 = 0.043\ m$
$O_1O_2 = L_1 = 0.202\ m$

**Link 2**
$O_2D_2 = d_6 = 0.185\ m$
$D_2A_2 = d_7 = 0.086\ m$
$O_2A_2 = d_8 = 0.204\ m$
$O_2B_2 = d_9 = 0.089\ m$
$B_2C_2 = d_{10} = 0.083\ m$
$O_2M_2 = R_2 = 0.054\ m$
$O_2O_3 = L_2 = 0.202\ m$

**Fig. 1.** Kinematic diagram and geometric dimensions of the humanoid's lower-limb system.

The moment inertia and total mass of each link are $I_1, m_1$ and $I_2, m_2$ respectively. The length of each SEA in every instant (noted as $L_{S1}, L_{S2}$) are functions of the respective rotation angles $\theta_1$ and $\theta_2$. By applying the *cosine rule* for triangles $O_1A_1B_1$ and $O_2A_2B_2$, the lengths of the SEAs are calculated as below (proven in [13]):

$$\leftrightarrow \begin{cases} L_{S1} = \sqrt{d_3^2 + d_4^2 - 2d_3 d_4 \cos\left(\frac{\pi}{2} - \sigma_1 + \theta_1 - \alpha_1\right)} \\ L_{S2} = \sqrt{d_8^2 + d_9^2 - 2d_8 d_9 \cos(\pi - \sigma_2 + \theta_2 - \alpha_2)} \\ L_{S1} = \sqrt{d_4^2 + d_3^2 - 2d_4[d_2 \cos(\theta_1 - \alpha_1) - d_1 \sin(\theta_1 - \alpha_1)]} \\ L_{S2} = \sqrt{d_8^2 + d_9^2 - 2d_9[d_7 \sin(\alpha_2 - \theta_2) - d_6 \cos(\alpha_2 - \theta_2)]} \end{cases} \quad (1)$$

The dynamical functions of the double compound pendulum system are determined by utilizing the Lagrange method. For the sake of simplicity, the following terms are defined:

$$\begin{aligned} \mathcal{M}_{11} &= m_1 R_1^2 + I_1 + m_2 L_1^2 + m_2 R_2^2 + 2 m_2 L_1 R_2 \cos\theta_2 \\ \mathcal{M}_{12} &= m_2 R_2^2 + m_2 L_1 R_2 \cos\theta_2 \\ \mathcal{M}_{21} &= m_2 R_2^2 + m_2 L_1 R_2 \cos\theta_2 \\ \mathcal{M}_{22} &= m_2 R_2^2 + I_2 \\ \mathcal{N}_1 &= -m_1 g R_1 \sin\theta_1 - m_2 g L_1 \sin\theta_1 + m_2 g R_2 \sin(\theta_2 - \theta_1) \\ &\quad - m_2 L_1 R_2 \sin\theta_2 \left(\dot\theta_2^2 - 2\dot\theta_1 \dot\theta_2\right) - B_1 \dot\theta_1 \end{aligned} \quad (2)$$

$$\mathcal{N}_2 = -m_2 g R_2 \sin(\theta_2 - \theta_1) - m_2 L_1 R_2 \dot{\theta}_1^2 \sin\theta_2 - B_2 \dot{\theta}_2$$

Then, the dynamical functions for the rotation angles of $\theta_1$ and $\theta_2$ are:

$$\begin{cases} \mathcal{M}_{11}\ddot{\theta}_1 - \mathcal{M}_{12}\ddot{\theta}_2 = \tau_{S1} + \mathcal{N}_1 + d_1 \\ \mathcal{M}_{22}\ddot{\theta}_2 - \mathcal{M}_{21}\ddot{\theta}_1 = \tau_{S2} + \mathcal{N}_2 + d_2 \end{cases} \quad (3)$$

In which, $\tau_{S1}, \tau_{S2}$ are the control torques created by the SEAs at the link $O_1 O_2$ and $O_2 O_3$; $B_1, B_2$ are the viscous coefficients at joint $O_1$ and $O_2$; and $d_1, d_2$ are the bounded disturbance torques acted about joint $O_1$ and $O_2$.

Furthermore, the torques generated by the SEAs can be calculated as:

$$\begin{cases} \tau_{S1} = -k_1 \Delta_1 d_4 \sin(\gamma_1) \\ \tau_{s2} = -k_2 \Delta_2 d_9 \sin(\gamma_2) \end{cases} \quad (4)$$

In which, $k_1$ and $k_2$ are the spring stiffness coefficients, $\Delta_1$ and $\Delta_2$ are the spring's deformations, $\gamma_1$ and $\gamma_2$ are varied angles (shown in **Fig. 1**) of respective joints.

By using Eq. (3) and Eq. (4), the dynamical equations are simplified as:

$$\begin{cases} \mathcal{M}_{11}\ddot{\theta}_1 - \mathcal{M}_{12}\ddot{\theta}_2 = -k\Delta_1 d_4 \sin(\gamma_1) + \mathcal{N}_1 + d_1 \\ \mathcal{M}_{22}\ddot{\theta}_2 - \mathcal{M}_{21}\ddot{\theta}_1 = -k\Delta_2 d_9 \sin(\gamma_2) + \mathcal{N}_2 + d_2 \end{cases} \quad (5)$$

### 2.2 SEA Modelling

By referring to our previous research in [13], the dynamical equations for the ball screw nut's motions inside the SEAs at two joints are given as:

$$\begin{cases} \ddot{r}_1 = U_{v1} - 48\dot{r}_1 \\ \ddot{r}_2 = U_{v2} - 48\dot{r}_2 \end{cases} \quad (6)$$

In which, $r_1, U_{v1}$ and $r_2, U_{v2}$ are the nut's displacement and input voltage of each joint, respectively.

Next, it is necessary to identify the dynamical characteristic of the end effector point of the SEA. Let $r_{B1}$ and $r_{B2}$ respectively be the linear displacements of the SEA's end effector at each joint. Define $\Delta_1 = r_{B1} - r_1$ and $\Delta_2 = r_{B2} - r_2$, the dynamical equation of the whole SEAs, with voltage input and internal spring deformation output is:

$$\begin{cases} \ddot{\Delta}_1 = \ddot{r}_{B1} + 48\dot{r}_1 - U_{v1} \\ \ddot{\Delta}_2 = \ddot{r}_{B2} + 48\dot{r}_2 - U_{v2} \end{cases} \quad (7)$$

Finally, to simplify the obtained dynamical equations, the following matrices are defined:

$$\boldsymbol{\theta} \triangleq \begin{bmatrix} \theta_1 \\ \theta_2 \end{bmatrix} \quad \boldsymbol{v}_x \triangleq \begin{bmatrix} \Delta_1 \\ \Delta_2 \end{bmatrix} \quad \boldsymbol{M} \triangleq \begin{bmatrix} \mathcal{M}_{11} & -\mathcal{M}_{12} \\ -\mathcal{M}_{21} & \mathcal{M}_{22} \end{bmatrix}$$

$$\boldsymbol{r} \triangleq \begin{bmatrix} r_1 \\ r_2 \end{bmatrix} \quad \boldsymbol{f}_X \triangleq \boldsymbol{M}^{-1} \begin{bmatrix} \mathcal{N}_1 \\ \mathcal{N}_2 \end{bmatrix} \quad \boldsymbol{g}_x \triangleq \boldsymbol{M}^{-1} \begin{bmatrix} -k_1 d_4 \sin(\gamma_1) & 0 \\ 0 & -k_2 d_9 \sin(\gamma_2) \end{bmatrix} \quad (8)$$

$$\boldsymbol{F}_R \triangleq \begin{bmatrix} \ddot{r}_{B1} \\ \ddot{r}_{B2} \end{bmatrix} \quad \boldsymbol{U}_v \triangleq \begin{bmatrix} U_{v1} \\ U_{v2} \end{bmatrix} \quad \boldsymbol{d} \triangleq \begin{bmatrix} \frac{-d_1}{k_1 d_4 \sin(\gamma_1)} \\ \frac{-d_2}{k_2 d_9 \sin(\gamma_2)} \end{bmatrix}$$

As a result, sets of Eq. (5) and Eq. (7) are rewritten in the matrix form as:

$$\begin{cases} \ddot{\boldsymbol{\theta}} = \boldsymbol{f}_x + \boldsymbol{g}_x(\boldsymbol{v}_x + \boldsymbol{d}) \\ \ddot{\boldsymbol{\Delta}} = \boldsymbol{F}_R + 48\dot{\boldsymbol{r}} - \boldsymbol{U}_v \end{cases} \quad (9)$$

## 3 Controller Design

First, let's define $x_1 \triangleq \theta$; $x_2 \triangleq \dot{\theta}$ as the state variables. The first dynamic equation in Eq. (9) can be then rewritten as:
$$\begin{cases} \dot{x}_1 = x_2 \\ \dot{x}_2 = f_x + g_x(v_x + d) \end{cases} \quad (10)$$

Next, define the following matrices:
$$X \triangleq [x_1 \quad x_2]^T \qquad X_d \triangleq [x_d \quad \dot{x}_d]^T \qquad \widetilde{X} \triangleq [\widetilde{x}_1 \quad \widetilde{x}_2]^T$$
$$\widetilde{x}_1 \triangleq x_1 - x_d \qquad \widetilde{x}_2 \triangleq x_2 - \dot{x}_d \qquad \varepsilon \triangleq [\widetilde{X} \quad X_d]^T \triangleq \begin{bmatrix} \widetilde{x}_1 & x_d \\ \widetilde{x}_2 & \dot{x}_d \end{bmatrix}^T \quad (11)$$

In which, $X$ is the state variables, $X_d$ is the desired angle and angular velocity trajectories, $\widetilde{X}$ is the tracking error, and $\varepsilon$ is an extended error matrix utilized in the later event-triggered control process.

### 3.1 Event-triggered SMC

Before implementing the controller design work, the following assumptions must be considered:

- *Assumption 1:* The time derivative of desired trajectories $\dot{x}_d$ is a function of $x_d$, which means $\dot{x}_d = f_d(x_d)$
- *Assumption 2:* The functions $f_x, g_x$, and $f_d$ are Lipschitz in a compact set.
- *Assumption 3:* The following functions are bounded by positive constants: $\|\widetilde{X}\| \leq r, \|X_d\| \leq h, \|d\| \leq d_0, \|g_x\| \leq g_0$, in which $r$ is a designed value.

Then, the sliding function vector is given as follow:
$$s = c\widetilde{x}_1 + \widetilde{x}_2 \quad (12)$$

In which, $c$ is a positive constant. The control matrix $v_x$ is chosen as follow:
$$v_x = g_x^{-1}[-f_x + \ddot{x}_d - c\widetilde{x}_2 - \rho \, \text{sign}(s)] \quad (13)$$

where $\rho$ is a constant that $\rho \geq g_0 d_0 + \eta$ $(\eta > 0)$. Next, the Candidate Lyapunov Function (CLF) for the first dynamical equation is proposed as below:
$$V_x = \frac{1}{2} s^T s \quad (14)$$

Taking the first-time derivative of Eq. (14), it yields:
$$\dot{V}_x = s^T[c\widetilde{x}_2 + f_x + g_x(v_x + d) - \ddot{x}_d] \leq -\eta\sqrt{1+c^2}\|\widetilde{X}\| \quad (15)$$

Let $t_i$ be the $i^{th}$ triggered instant, $t$ is the period between two consecutive triggered instants, which means $t \in [t_i; t_{i+1})$. The difference between $\varepsilon(t_i)$ at a triggered instant and $\varepsilon(t)$ are called the event-triggered error:
$$e(t) = \varepsilon(t_i) - \varepsilon(t) \quad (16)$$

Consequently, the control law at the triggered instant is inferred as:
$$v_x(t_i) = g_x^{-1}(t_i)[-f_x(t_i) + \ddot{x}_d(t_i) - c\widetilde{x}_2(t_i) - \rho \, \text{sign}(s(t_i))] \quad (17)$$

***Noted:*** *For simplified expression, the afterwards value of variables at the feedback sampling instant $t$ would be kept in original form (e.g. $x_1$ instead of $x_1(t)$).*

As the event-triggered law now controls the lower-limb system in Eq. (17), the first derivative of the CLF along the trajectories of Eq. (13) is rewritten as:

$$\dot{V}_x = \frac{\partial V_x}{\partial \widetilde{X}} \frac{d\widetilde{X}}{dt} = \frac{\partial V_x}{\partial \widetilde{X}} \begin{bmatrix} \dot{\widetilde{x}}_1 \\ f_x + g_x(v_x + d) - \ddot{x}_d \end{bmatrix} + \frac{\partial V_x}{\partial \widetilde{X}} \begin{bmatrix} 0 \\ g_x(v_x(t_i) - v_x) \end{bmatrix}$$
$$\leq -\eta\sqrt{1+c^2}\|\widetilde{X}\| + (1+c^2)\|\widetilde{X}\|L\|e\| \qquad (L>0) \tag{18}$$

Consequently, the system is ensured to be stable when the following that is called the event-triggered condition, holds:
$$\|e\| < \frac{\eta}{L\sqrt{1+c^2}} \tag{19}$$

Since the inter-event time $T_i$ features the Zeno phenomenon, it is essential to determine the lower bound of $T_i$. Considering the first time-derivative of tracking error matrix, it yields:
$$\dot{\widetilde{X}} = \dot{X} - \dot{X}_d = \begin{bmatrix} x_2 \\ f_x \end{bmatrix} + \begin{bmatrix} 0 \\ g_x \end{bmatrix}(v_x + d) - \begin{bmatrix} \dot{x}_d \\ \ddot{x}_d \end{bmatrix} \tag{20}$$

Taking the $2-norm$ of Eq. (20), it can be obtained:
$$\|\dot{\widetilde{X}}\| = \left\| \begin{bmatrix} x_2 \\ f_x \end{bmatrix} + \begin{bmatrix} 0 \\ g_x \end{bmatrix}(v_x(t_i) + d) - \begin{bmatrix} \dot{x}_d \\ \ddot{x}_d \end{bmatrix} \right\|$$
$$\leq \lambda(r + h + d_0 + \|e\|) \qquad (\lambda > 0) \tag{21}$$

From the relations defined in Eq. (16) and Eq. (21), the first time-derivative of event-triggered error matrix $e$ can be calculated as:
$$\frac{d\|e\|}{dt} \leq \|\dot{e}\| \leq \|\dot{\widetilde{X}}\| \leq \lambda(r + h + d_0 + \|e\|) \tag{22}$$

Consequently, from differentiating inequation in Eq. (22), the inter-event time lower bound is determined as:
$$T_i \geq \frac{1}{\lambda} \ln\left(r + h + d_0 + \frac{\eta}{L\sqrt{1+c^2}}\right) > 0 \tag{23}$$

### 3.2 Backstepping process

Consider the first augmented dynamic system:
$$\begin{cases} \dot{X} = \begin{bmatrix} x_2 \\ f_x + g_x d \end{bmatrix} + \begin{bmatrix} 0 \\ g_x \end{bmatrix} z_1 \\ \dot{z}_1 = v_1 \end{cases} \tag{24}$$

As the control law $v_x(t_i)$ is proven to be able to ensure the stability of the overall system, the control goal now is to drive $z_1$ follow the desired value of $v_x(t_i)$. First, the augmented CLF is proposed as:
$$V_1 = V_x + \frac{1}{2}(v_x(t_i) - z_1)^T(v_x(t_i) - z_1) \tag{25}$$

The first time-derivative of $V_1$ yields:
$$\dot{V}_1 = s^T[c\widetilde{x}_2 + f_x + g_x(v_x(t_i) + d) - \ddot{x}_d]$$
$$+ s^T g_x(z_1 - v_x(t_i)) + (v_x(t_i) - z_1)^T(\dot{v}_x(t_i) - v_1) \tag{26}$$

To stabilize the proposed extended system, the first pseudo-control law is:
$$v_1 = k_{p1}(v_x(t_i) - z_1) + \dot{v}_x(t_i) - (g_x)^T s \qquad (k_{p1} > 0) \tag{27}$$

Through the backstepping design method, the second extended system is proposed as:

$$\begin{cases} \dot{X}_1 = f_1(X_1) + g_1(X_1)z_2 \\ \dot{z}_2 = f_2(X_2) + g_2(X_2)U_v \end{cases} \quad (28)$$

where: $X_1 \triangleq \begin{bmatrix} X \\ z_1 \end{bmatrix}$, $f_1(X_1) \triangleq \begin{bmatrix} x_2 \\ f_x + g_x d \\ 0 \end{bmatrix} + \begin{bmatrix} 0 \\ g_x \\ 1 \end{bmatrix} z_1$, $g_1(X_1) \triangleq \begin{bmatrix} 0 \\ 1 \end{bmatrix}$, $X_2 \triangleq \begin{bmatrix} X_1 \\ z_2 \end{bmatrix}$,

$f_2(X_2) \triangleq F_R + 48\dot{r}$, $g_2(X_2) \triangleq -1$

The $X_1$-system is assumed to be stable by the control law $v_1$. Then, to make $z_2$ reach the value of $v_1$, the $2^{nd}$ augmented CLF is proposed as:

$$V_2 = V_1 + \frac{1}{2}(v_1 - z_2)^T(v_1 - z_2) \quad (29)$$

Take the first time-derivative of Eq. (29), it yields:

$$\begin{aligned}\dot{V}_2 &= \dot{V}_x + (v_x(t_i) - z_1)^T(\dot{v}_x(t_i) - v_1) + (v_x(t_i) - z_1)^T(v_1 - z_2) \\ &\quad + (v_1 - z_2)^T[\dot{v}_1 - f_2 - g_2 U_v]\end{aligned} \quad (30)$$

In order to prove that the error term $(v_1 - z_2)$ is bounded, the first time-derivative of $V_2$ must be proven to be $\leq 0$. Then, the control input signal of the second extended system, which is also the input voltage of the SEA, is chosen as:

$$U_v = g_2^{-1}[-f_2 + k_{p2}(v_1 - z_2) + \dot{v}_1 + (v_x(t_i) - z_1)] \quad (k_{p2} > 0) \quad (31)$$

## 4　Simulation and Analysis

In this section, a simulation is conducted to compare the response quality between traditional SMC and event-triggered SMC. Generally, the remote-control process for the robotics system suffers significant delay due to massive data transmission and computer resource consumption. Besides, the low sampling time control signal might become redundant since the heavy physical structure of the robot can hardly respond quickly enough to every single updated demand. Therefore, the event-triggered SMC would be applied to deal with the above downsides.

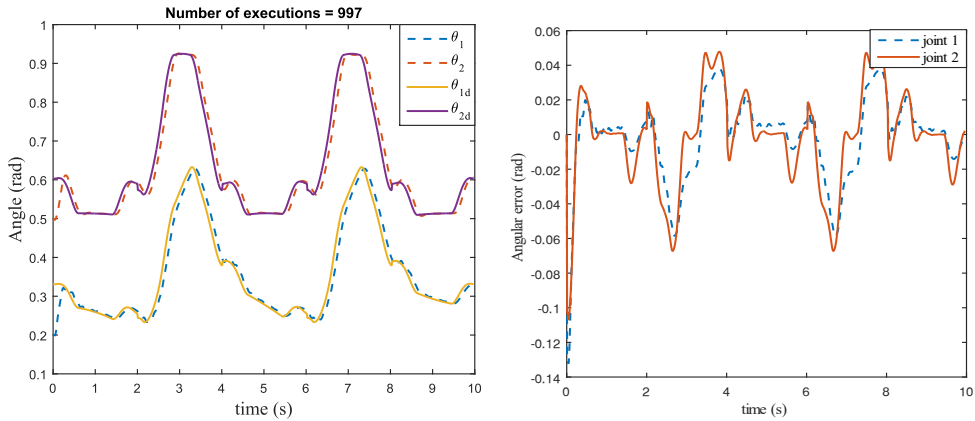

**Fig. 2.** The rotational angles response of the lower-limb system (left) and the tracking errors (right) when being controlled by traditional SMC

The desired trajectories of rotation angle are derived from our previous research in [13]. The period of one walking cycle in simulation is $10\ s$ and the sampling time of output state is $10\ ms$. The physical parameters of the lower-limb system are: $m_1 = 1.07\ kg$, $m_2 = 0.89\ kg$, $I_1 = 0.028\ kgm^2$, $I_2 = 0.002\ kgm^2$, $K_{s1} = 20000\ N/m$, $K_{s2} = 20000\ N/m$, $B_1 = 0.3\ Nms/rad$, $B_2 = 0.3\ Nms/rad$.

First, the lower-limb system is controlled with the traditional SMC strategy. A simple tuning process is applied to find out the appropriate controller's gain values, which is $c = 30, \rho = 20, k_{p1} = 3, k_{p2} = 5$. The resulted responses and tracking errors are demonstrated in **Fig. 2**.

Overall, the tracking process shows a constant delay compared to the desired trajectories due to the inertia of the lower-limb system and the integration of the backstepping process. The maximum error at both joints is about $0.07\ rad$ and mainly occurs at the peaks of the graphs where the links suddenly change their rotational direction. The number of execution instants triggered by traditional SMC is 997, which is almost the same as the number of output samples taken in one walking cycle.

Next, an event-triggered SMC strategy is applied for the humanoid's lower-limb system. The controller gain values utilized in the traditional SMC are kept unchanged. The control law of the event-triggered SMC is designed with the bounding coefficients of: $r = 0.2, \eta = 0.3, L = 0.85$.

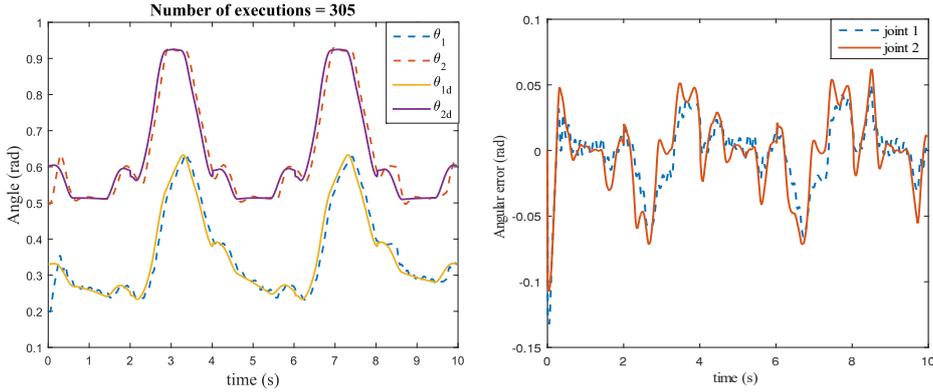

**Fig. 3.** The angles response of the lower-limb system (left) and the tracking errors (right) when being controlled by event-triggered SMC

The results illustrated in **Fig. 3** prove the almost equal quality of response from the event-triggered SMC compared to the traditional one. In specific, besides the delay characteristic from the system itself, the tracking errors in the event-triggered SMC case are slightly larger, about $0.075\ rad$ at the peak of the graph. However, in 90% of the other part of the graph, the tracking errors of both control strategies are the same. Moreover, it should be noted that the number of executions in event-triggered SMC, with only 305 triggered instants, is significantly lower than that of traditional strategy. This number clearly shows the advantage of event-triggered SMC when it can help reduce the total transmission time by more than threefold, which saves up much more computer resources than the traditional SMC.

As shown in **Fig. 4**, the highest inter-event time can be up to $0.25\ s$ and mainly maintains at about $0.05\ s$, which proves for the efficiency of the proposed controller, compared to the constant $0.01\ s$ inter-event time of traditional SMC. This means the controller only generates and sends the new control signal every $0.05\ s$ in average while the sampling time in the whole system is $0.01\ s$, which significantly reduces the load of data needed to transmit in the entire process and considerably save up the calculation time. In the periods when the control signal is kept unchanged, the lower-limb system is ensured to maintain the tracking process with the tracking error $\|\tilde{X}\|$ lower than the designed value $r$. The trigger instant occurs whenever $\|\tilde{X}\| > r$ or when the Lyapunov stability standard of the system is violated.

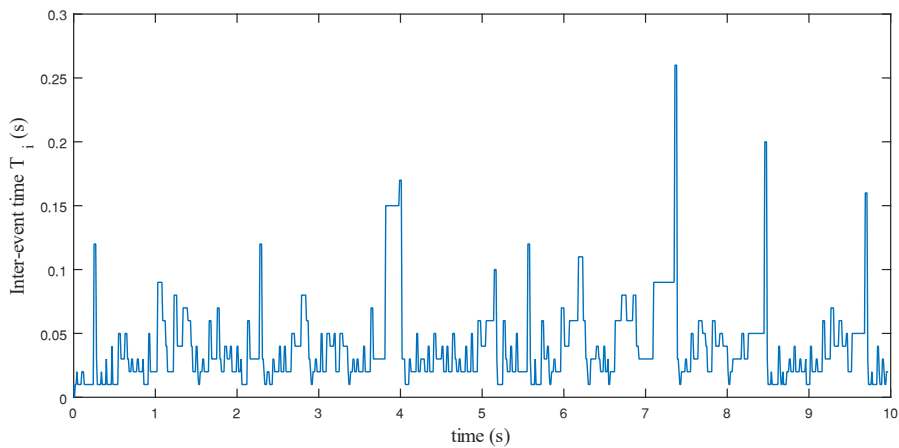

**Fig. 4.** The inter-event time of the humanoid's lower-limb system when being controlled by traditional SMC

## 5 Conclusion

In summary, the event-triggered SMC strategy has clearly shown its advantages when applied in the complex physical system with considerable inertia like lower-limb powered by SEAs. With the remarkable potential of reducing overall execution time, saving up resources for transmission, and consuming less energy during operation, the event-triggered approach in general and event-triggered SMC, in particular, would be appropriate for practical applications. In the future, a comprehensive practical experiment will be conducted to further evaluate the efficiency and optimize the performance of this control strategy.

For further research intention, the Event-triggered SMC algorithm will be utilized to control the whole lower body of the humanoid robot with 6 DOFs powered by the SEAs during a walking cycle. The entire process would be controlled and managed remotely between the humanoid and the master control device through a wireless transmission protocol, which may lead to some delay phenomenon as for the distant control and for the high complexity of the system itself. In that case, the event-triggered SMC would be an appropriate choice to optimize the amount of data transmitted to the humanoid and considerably save up consumed energy.

## Acknowledgements

This research is supported by DCSELAB and funded by Vietnam National University Ho Chi Minh City (VNU-HCM) under grant number TX2022-20b-01. We acknowledge the support of time and facilities from Ho Chi Minh University of Technology (HCMUT), VNU-HCM for this study.